# EZYer: A simulacrum of high school with generative agent

Jinming Yang[†]
University of Electronic Science
and Technology of China
Chengdu, China
202522080607@std.uestc.edu.cn

Zimu Ji
Shaanxi University of Science
and Technology
Xi'an , China
202215030214@sust.edu.cn

Weiqi Luo
Shaanxi University of Science
and Technology
Xi'an , China
202315030214@sust.edu.cn

Gaoxi Wang
Shaanxi University of Science
and Technology
Xi'an , China
202315020203@sust.edu.cn

Bin Ma
Shaanxi University of Science
and Technology
Xi'an , China
202315030113@sust.edu.cn

Yueling Deng
Shaanxi University of Science
and Technology
Xi'an , China
202315030202@sust.edu.cn

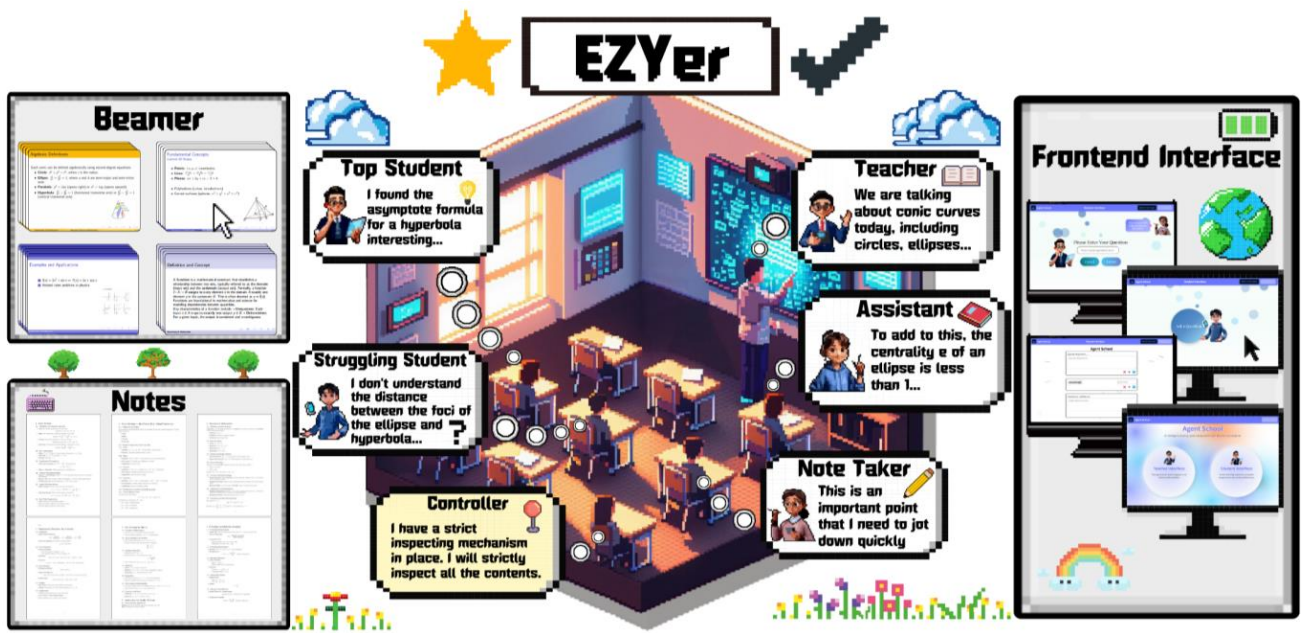

Figure 1: Concept map of EZYer. The concept map visualizes the interactive interface, generative objects, and three functions of EZYer. First, to generate a Beamer that can be used in the high school math lesson based on the problems entered and materials and images uploaded by the user. Second, to design five generative agents based on the main roles in the high school math lesson. Each of these agents embodies a different personality and function , and interacts with each other according to elaborate prompts and their relationships, ultimately generating a well-formatted and well-organized Notes. Third, we set up an inspecting mechanism to ensure that all key features run smoothly through multiple dimensions of content inspection. With EZYer, we would like to explore the potential of LLM-empowered generative agents to simulating actual human behavior in instructional scenarios.

## ABSTRACT

With the rapid development of the online education and large language model, the existing educational tools still suffer from incomplete service, insufficient performance and weak interactivity in terms of courseware generation, interactive notes and quality assurance of content. In particular, the proposed generative agent EZYer：1) Teacher Module: Integrating the Text Corpus retrieval and in-depth generation technologies, it automatically generates structured teaching materials and LaTeX Beamer courseware in line with the high school mathematics syllabus and supports user-defined image insertion. 2) Student Module: Throughout the collaborative interaction of the four roles of Teacher, Assistant, Top Student and Struggling Student, Note Taker summarizes and generates academic notes to enhance the depth and interest of learning. 3) Controller: set up keyword filtering system, content scoring system, role co-validation system, and dynamic content correction system. This ensure academic strictness and pedagogical propriety of EZYer inputs and outputs. In order to evaluate EZYer, this paper designs five-dimensional evaluation indexes of content accuracy, knowledge coverage, usability, formatting correctness and visual design and appeal, and scores 100 Beamer and Notes generated by EZYer by five large language models, separately, and the results show that the quality of EZYer-generated content is excellent and has a good application prospect.

## 1 INTRODUCTION

In the current education model, teachers always need to integrate teaching materials before class and create clear courseware to assist their teaching [1], while students need to summarize the knowledge points of the day after class for review and learning [2]. Recently, AI Agent has made excellent breakthroughs in the field of education [3, 4, 5, 6, 7], and according to the feedback from teachers and students who have used educational tools [8, 9, 10] up until now, the currently released tools generally contract the shortages of incomplete service, insufficient performance, weak interactivity, and supplying of misleading information [11, 12] for instance, none of them have the ability to generate LaTeX Beamer functionality.

Therefore, in consideration of fill the gap of Educational Agents in this area, we design EZYer: a generative agent of mathematical knowledge for high school teachers and students, where we choose the DeepSeek-V3 model. The EZYer was consisted from three core units: Teacher Module, Student Module, and Controller, which is established the generative agents [17, 18] of the large language model [13, 14, 15, 16]. After coming in Teacher Module, users only need to type the education content in the input box or upload a PDF/Word document. Firstly EZYer will detect whether the materials is consistently with Text Corpus automatically [19, 20] and if it matches, it will directly call the relevant content from Text Corpus and generate the teaching content and exercises. If it does not match, it will call DeepSeek API to generate the education materials and exercises, and generate LaTeX Beamer based on the teaching materials [21]. Not only that, teachers can also insert images into the Beamer according to their own ideas. We design different AI characters in Student Module to deeply replace the interactive learning scenarios [24, 25, 26, 27, 28, 29] of a real hall [22, 23], so that the students can experience the interactive learning style [30, 31] after entering Student Module. By simply typing their questions in the input box, the material is also automatically detected firstly whether it counterparts the content in Text Corpus. If it matches, the answer to the question is generated directly based on the set prompt and the related content in Text Corpus. If it does not match, DeepSeek API will be called to set up the content of the dialog. After passing the test, each role (Teacher, Assistant, Top Student, Struggling Student) will start outputting knowledge points, supplementary content, expanded content and error-prone points in proper sequence. At the end of the discussion, another role (Note Taker) will summarize the conversation and generate Notes in PDF format. At the same time, in order to insure that EZYer can provide a good user experience all along operation, we have designed a inspecting mechanism (Controller) inside EZYer to review each time when users enter content, and to point out and block irrelevant content in a timely style, so as to provide the validity of the output content and to enhance the user experience of teachers and students.

The characteristic of EZYer lies in:

(1) Teacher Module generates a professional Beamer that obeys the academic publishing standards based on an uploaded document or a simple description of the input.

(2) Student Module has interactive facial characteristics and could create Notes by summarizing the output of all roles.

(3) Controller makes sure the validity of the input and output content.

In order to systematically evaluate the reliability of our design of EZYer, we evaluated the generators (Beamer and Notes) of Teacher Module and Student Module using unique large language models. By setting different scoring mechanisms for Beamer and Notes respectively, and scoring them using five different large language models [32, 33], positive evaluation results were obtained, which proved that the generators of EZYer have good application prospects.

Technological contributions of EZYer:

(1) As a professional and effective teaching tool, it raises the efficiency of teachers' lesson arrangement and the quality of instructional content through the innovative application of AI-generated content, and also guarantees that teachers in any technological background are able to make artistic education materials and devote more energy to teaching itself.

(2) Based on interactive learning and personalized summaries, it provides students with diverse learning perspectives to help them understand the knowledge points from various levels and depths, that greatly helps the personalization and efficiency of study.

(3) The design of Controller avoids the interference of invalid information, ensures that the input content is always in accordance with the teaching and learning aims, and effectively boosts the stability and dependability of the system.

In addition to the above technological innovations, EZYer has further made outstanding contributions in terms of educational equity and personalized learning support. (1) It reduces the pressure of teachers in education resource-poor regions on curriculum design and helps the utilization efficiency of educational resources, so promoting a more equitable distribution of instructional resources. (2) It can significantly enhance the efficiency and quality of education with the student public in underdeveloped areas. (3) EZYer breaks the geographic limitations and is able to be regionally adjusted according to the educational needs of different nations and regions, so that it can serve the worldwide users. This combination of globalization and localization will supply powerful support for educators all around the world.

## 2 EZYER

### 2.1 Teacher Module

In order to help our users own a better experience when generating their ideal Beamer, EZYer sets up a framework to exchange the data flow, and make sure there is a clear division of responsibilities. Upon different stages, different work there is.

The first work in Teacher Module is the enter content, to input and upload PDF/Word documents. Secondly, chapters can be split well enough in a clear way when processiong, and EZYer will detect whether the content is already recorded in Text Corpus. If the counterparts can be matched, EZYer will use Text Corpus to generate content. On the other hand, EZYer will call DeepSeek API for the generation of teaching materials and exercises. Moreover, having pictures, clear type setting, and correct math formula as the essential parts of a presentation can be extremely important and indispensable. That is why the last step of output is matter. Users upload and insert the pictures follow their ideas, then EZYer will escape the providing materials, exercises and pictures into LaTeX, and compile it into Beamer document which is formatted by PDF.

This clear and complete architecture method defines the roles of each part, from input to processing and output. It ensures a smooth data flow, escaping the original text input and turning it into a final PDF.

Since we have mentioned the three duties of Teacher Module, and simply list part of their functions, we are going to introduce how Teacher Module achieved them in detail and logical words.

The users first enter EZYer, type the knowledge points or keywords that need to be generated into LaTeX Beamer. Teacher Module uses the multi-modal input stream and paragraph semantic boundary detection to intelligently segment the unstructured raw text into a collection of knowledge point units, and searches for the existence of the knowledge point units in the collection in Text Corpus, and if it is contained in Text Corpus, it will directly called Text Corpus and generates teaching materials and exercises. If it does not exist in Text Corpus, the big model is called to perform a bidirectional transformation: the simplified knowledge point units are expanded into structured teaching materials.

We designed a set of structured prompt, templating to guide the large language model in generating teaching materials and exercises, and the prompt for the generated materials sections are provided here as examples:

**Prompt**
{
Transform the following raw content (which may be very brief ) into a detailed, structured academic teaching material in English.
The generated content must strictly adhere to the high school mathematics curriculum and teaching standards.

Please provide a comprehensive explanation that includes:
- A clear definition of the topic.
- An in-depth explanation covering fundamental concepts.
- Mathematical formulations (in LaTeX notation) where applicable.
……
Format requirements:
-Use Markdown with section headers (## for chapter titles).
-Present mathematical formulas in LaTeX notation.
-Organize the content using bullet points and paragraphs.
-The output must be strictly in English and written in a formal academic tone.
……
}

The prompt for generating exercises are similar to those for generating instructional materials. Please refer to Appendix A for the whole prompt.

The use of structured prompts significantly improves the quality of the instructional materials and exercises that generated by the large language model, and strictly ensure the generated mathematical formulas are in LaTeX format. So the users do not have to manually enter complex mathematical formulas, which greatly saves the time and improves the convenience. Finally, the user interface generates formatted Markdown teaching materials and exercises for reading.

Once this two operations are completed, Teacher Module can receives all the generated teaching materials. Then the user will be able to upload images for insertion into the selected page.

We designed another set of prompts (turn to Appendix A) to make sure the generated LaTeX code obey the formation of Beamer. So EZYer can handle the image path and insertion logic.

We transform the Markdown syntax and exercises into LaTeX code, and escape the special characters in the picture path into LaTeX. Then, we return the LaTeX code which can successfully compile on the website. Finally, compile these code into LaTeX Beamer formatted by PDF, so users can download it.

Instead of simply presenting the generated instructional materials and exercises in Beamer directly, we re-partitioned the materials and designed a series of strict prompt for the large language model, so we could generate the content more streamlined and briefly presented in Beamer.

In order to systematically show the functions and algorithmic logic of Teacher Module, we select the method of formal mathematical model, through a series of mathematical formulas to define the content processing segmentation mechanism, teaching materials and image path escaping, compilation mechanism, LaTeX compilation mechanism, in order to reveal its inner principle and logic:

$$\mathcal{P}(I,\theta,\gamma) = \mathcal{C} \circ \mathcal{B}\big((\mathcal{M} \otimes \mathcal{E}) \circ \mathcal{S}(I),\theta,\gamma\big) \quad (1)$$

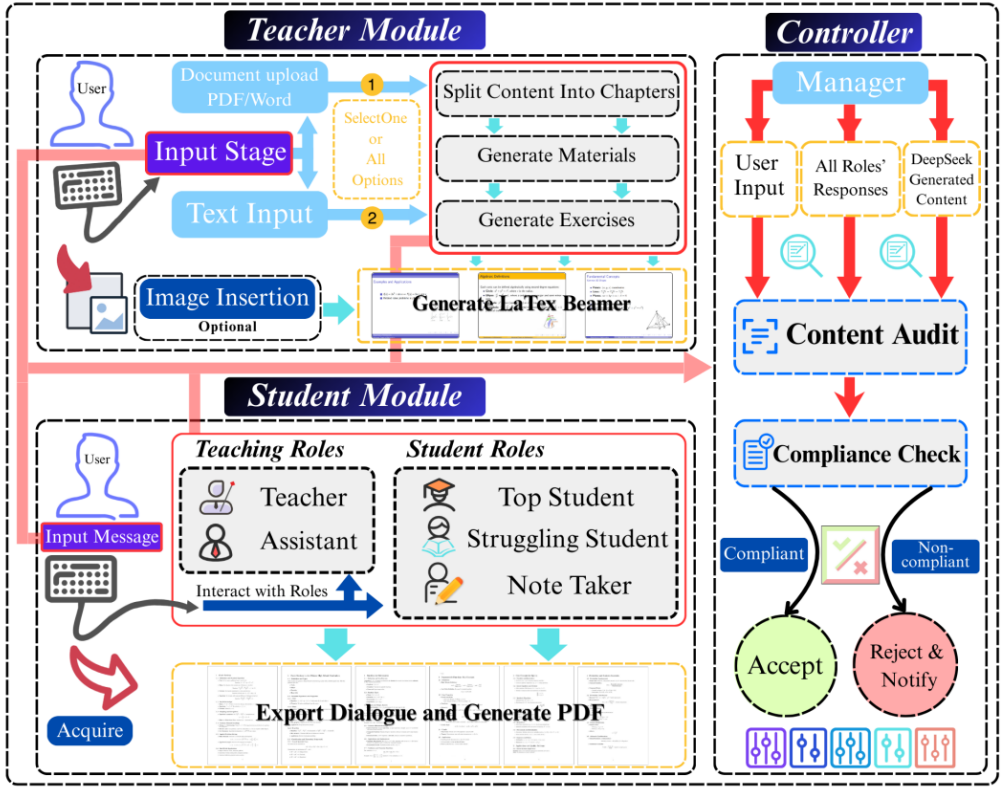

Figure 2: The overall framework of EZYer. EZYer consists of three core components: Teacher Module, Student Module, and Controller. Teacher Module generates materials based on the text and questions uploaded by the user, parses them, and then summarizes and generates a beamer that can be used for teaching and learning, according to the generated materials and the inserted images. Based on the generated materials and inserted images, it summarizes and generates a beamer that can be used for teaching, while Student Module matches the user's input with Text Corpus and calls the API of the large language model to realize the interaction of the five roles, and finally can export a high-quality notes with a clear structure and multi-faceted coverage. On the other hand, Controller verifies the user's input and the intellectual output to ensure the accuracy of incoming and transmitted material.

Where I is the user input, $\mathcal{S}$ is defined as the chapter segmentation function, and we set a parallel processing operator $\otimes$ to generate teaching materials and exercises in parallel for the content processed by $\mathcal{S}$ through the teaching material generation function $\mathcal{M}$ and the exercises generation function $\mathcal{E}$. Here we also define $\theta$ as the user uploads the picture parameters of the picture, uploaded successfully set $\gamma$ used to indicate the picture location parameters, LaTeX Beamer generator function $\mathcal{B}$, respectively, $\mathcal{M}$, $\mathcal{E}$, $\theta$, $\gamma$ in the content and parameters of the compilation, and finally through the compiler function $\mathcal{C}$, compiled to PDF format.

In order to be able to better understand the equation (1), we will decompose the formula, each decomposition of the formula are explained.

$$\mathcal{S}(I) = \{\mathcal{C}_1, \mathcal{C}_2, \ldots, \mathcal{C}_n\} \quad (2)$$

Equation (2) uses the $\mathcal{S}$ function to partition the contents of user input I into n sections to facilitate subsequent operations.

$$(\mathcal{M} \otimes \mathcal{E}) \circ \mathcal{S}(I) = \left((\mathcal{M}_1, \{\mathcal{E}_{1j}\}), (\mathcal{M}_2, \{\mathcal{E}_{2j}\}), \ldots, (\mathcal{M}_n, \{\mathcal{E}_{nj}\})\right) \quad (3)$$

Equation (3) uses $\otimes$ to parallelize the content of the segmented chapters to generate instructional materials and practice problems via the $\mathcal{M}$ and $\mathcal{E}$ functions.

$$\mathcal{B}(\{(M_i, \{\mathcal{E}_{ij}\})\}, \theta, \gamma) = L \quad (4)$$

Equation (4) mainly integrates the content generated by M function and E-function as well as the $\theta$, $\gamma$ of user uploaded images into LaTeX Beamer through $\mathcal{B}$ function.

$$\mathcal{C}(L) = P \quad (5)$$

Finally, Equation (5) compiles the integrated LaTeX Beamer to PDF using the $\mathcal{C}$ function.

The whole process fully considered the elements required for presentation, such as images, typography, mathematical formulas,

accuracy and professionalism. To meet the needs of teaching scenarios in all aspects, especially at the end of the generation of Beamer in LaTeX format, EZYer allows users to get a more standardized and academic presentation, which is one of the main purposes of our design of EZYer.

## 2.2 Student Module

In EZYer, we suppose an interaction framework based on the role division of labor. In order to realistically simulate the interaction behaviors in the classroom, we designed five roles and carefully designed the prompt for each of the role, including Teacher, Assistant, Top Student, Struggling Student, and Note Taker.

Specifically, we selected a recursive and cumulative functional relationship to describe the interaction between the following roles: Teacher first generates the content, Assistant adds to it, Top Student deepens the content, Struggling Student raises the error-prone points, and Note Taker summarizes all the content and generates Notes. This process can be expressed as the following recursive cumulative formula:

$$T = f_0(I_0) \tag{6}$$

$$A = f_1(T, I_1) \tag{7}$$

$$TS = f_2(T + A, I_2) \tag{8}$$

$$SS = f_3(T + A + TS, I_3) \tag{9}$$

$$NT = f_4(T + A + TS + SS, I_4) \tag{10}$$

Where, T is the output of Teacher, A is the output of Assistant, TS is the output of Student, SS is the output of Struggling Student, and NT is the Note Taker's notes summarizing the conversation of the previous four roles. The output of each role expands on the output of all previous roles to generate more comprehensive and richer learning content.

The design of the role is shown in Figure 3:

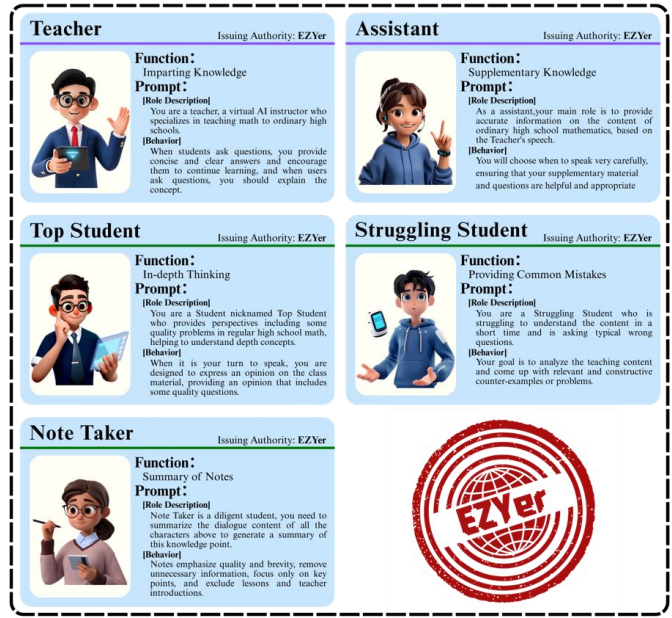

Figure 3: The "ID card" for each role. Each "ID card" contains the Name, Function and Prompt of the character

**Teacher:** Focuses on a systematic explanation of math concepts, presenting core knowledge points in a concise, clear and logical manner. Responds in a straightforward style and encourages students to explore further.
**Assistant:** Builds on Teacher's explanations and provides additional details to help students strengthen their understanding. Responses are strictly centered on the core of Teacher explanation.
**Top Student:** Extends Teacher and Assistant explanations by asking more deeper questions. Top Student not only asks more complex questions based on Teacher and Assistant feedback, but also guides other students to think deeply.

**Struggling Student:** Struggling Student plays the role of a typical error-prone student in the class, providing feedback on common mistakes and difficulties in learning. The input of Struggling Student is based on the output of the previous roles, and reinforces the correct knowledge points through reverse thinking.
**Note Taker:** Note Taker is responsible for summarizing the main points of classroom discussions and generating summary notes in an academic format. Note Taker summarizes the interactions of the other roles and generates concise and polished academic notes for the students.

An example of the role prompt is provided here:
**Prompt**

```
{
   # Teacher

   Role("Teacher", "Teacher", "[Role Description] You are a teacher, a virtual AI instructor who specializes in teaching math to ordinary high schools.\n[Behavior] When students ask questions, you provide concise and clear answers and encourage them to continue learning, and when users ask questions, you should explain the concept.")
}
```

Please refer to Appendix B respectively for the whole prompt.

In order to enhance the realism of EZYer, the behaviors and interactions of both the teacher role and the student role have been carefully designed. This approach is not limited to simulating a single teaching process, but explains and summarizes knowledge points from multiple dimensions.

When the user enters the question, the keywords are first matched in Text Corpus, and if it is match successfully, Teacher takes the leading role in responding the question which is posed and based on users' own prompt and the information in Text Corpus. Teacher teaches the subject in the classroom to help the students set up a solid academic foundation. Therefore, Teacher output relies heavily on the definitions and standard content of Text Corpus, which usually includes basic conceptual explanations, definitions, examples and so on. So EZYer can make sure that students have an initial understanding of the problem and encourage them keep on learning. If the match fails, the current module will calls DeepSeek API interface to generate a response based on the context of the question itself and the default template set within the program.

Assistant follows the answer of Teacher with additions and extensions content. To help students understand the content deeply. Assistant provides more specific details, examples, or additional explanations based on the answer of Teacher and related concepts in Text Corpus. It allows students to build on their initial understanding and gain more additional knowledge.

This is followed by Top Student response, a role set up to push students to think further. Top Student will not only rely on the responses of Teacher and Assistant, but will also generate deeper thinking based on the relevant content in Text Corpus, which will inspire the student set up a solid understanding of the concept.

Then Struggling Student will bring up common student misconceptions or misinterpretations of knowledge points based on the content generated by Teacher, Assistant, and Top Student, thinking critically, asking counter-examples or questions to help other students correct their misunderstandings, and draw their attention to possible misconceptions.

Finally, Note Taker, acts as a diligent student in the classroom, listening attentively and organizing notes and formulas. Note Taker does not make any statements, but only organizes the results of the interactions of the other four roles. Note Taker retains the useful information from the conversations of the other four roles, filters Markdown formatting, and generates a well-structured, well-organized Notes document which is following the rules of LaTeX compliant syntax specification.

This well-designed role interaction, by simulating the real classroom environment, makes the learning experience more dimensional to our real world. Not only to experience the multi-dimensional knowledge points explained, but also to generate a high-quality PDF format notes at the end.

## 2.3 Controller

In EZYer, the Controller plays a special role in managing the requests from users, invoking the services, and handling the entire process, all along the process of input to the final output. Benefit from the efficient operation of Controller, EZYer can make sure that each link from the original text to the final generated teaching material, meets the predefined teaching goals and academic standards. Obviously, content review is the core to guarantee the operation of all the key functions mentioned above, so we set up an inspecting mechanism to ensure that the output of each link meets the academic standards and pedagogical appropriateness by reviewing the content in multiple dimensions, standards and pedagogical appropriateness. It consists of four systems: keyword filtering system, content scoring system, role co-validation system, and dynamic content correction system.

**Keyword Filtering System.** Controller first performs a preliminary check of the input content through the built-in keyword filtering mechanism. It detects whether the content contains valid keywords (e.g., "function", "probability", etc.) or disabled keywords (e.g., "differential equation", "linear algebra", etc.) in real time through regular expressions. The presence of valid keywords insure the content conforms to the high school mathematics syllabus. Meanwhile, disabled keywords do well in filter out those content which is not appropriate for high school education. This filtering mechanism is the basic content review step that guarantees the given topics and difficulty levels of the generated content are as expected.

**Content scoring system.** After completing the basic keyword filtering, Controller will further calls DeepSeek API to assess the quality of the content through the content scoring system. The scoring system based on a scale of 1-5, and the evaluation of dimensions include logical coherence, pedagogical suitability and content relevance. The system focuses on checking whether the generated content meets the requirements of the high school mathematics curriculum standards. The scoring system supplies Controller with an automated way of controlling the quality of the content, ensuring that each generation is not only linguistically standardized, but also pedagogically appropriate.

Examples of prompts for the scoring system are provided here:
**Prompt**
```
{
   You are a virtual AI education assistant. Evaluate the following response based on the content provided:
      {reply_text}
      Please rate the response from 1 to 5:
      1: Totally illogical or does not align with high school math curriculum.
      2: Incoherent or acknowledging the response is part of a game.
```

3: Not ideal but acceptable, contains irrelevant content.
4: Generally appropriate, contains minor awkwardness.
5: Perfectly appropriate, fits well with the context and educational goals of high school math.
Please provide a score with a brief reason.
}

**Role Collaboration Validation System.** Once the content passed the scoring system, Controller will enters the role validation phase. This is where Student Module defines four roles that can speak: Teacher, Teaching Assistant, Top Student, and Struggling Student, each of which is responsible for interacting with and validating the generated content from different perspectives. The output of each role is expanded based on the content of all previous roles: from the explanation of basic concepts, to the validation of in-depth thinking, to the reverse checking of error-prone points, each of them adds validation and depth to the final generated academic material. All people's responses are collected and reviewed as a whole by Controller. When the output score of any role falls below the score 3, Controller will automatically trigger the content correction process, requiring the re-generation of content that meets the criteria.

**Dynamic Content Correction System.** If content violations (such as low content score or keyword violation) are found during content scoring or role co-verification, Controller will automatically trigger dynamic content correction. For keyword violations, the system will return the specific offending terms and provide modification suggestions: if the content score is too low, API will be re-called to generate compliant content. For structural issues, Controller will analyze LaTeX compilation logs and make the necessary formatting corrections. This system ensures that content is automatically optimized and adjusted at every stage, ultimately ensuring that the generated teaching materials meet high academic standards.

## 3 EVALUATION

In this section, we evaluate the generation from users (Beamer and Notes) in processing of using EZYer, and to explore the potential of EZYer. The empowered by the large language model, shows in terms of the generative agent simulating the behaviors of high school teachers and students in a real teaching environment. For this purpose, we set up a specific scoring mechanism for evaluating the quality of its generated Beamer and Notes. The evaluation process will be accomplished by evaluating multiple materials that generated by several large language models for EZYer, for which it was decided to use five dimensions for a comprehensive evaluation of the generated documents: content accuracy, knowledge coverage, usability, formatting correctness, and visual design and appeal.

### 3.1 Scoring mechanism

In order to ensure that Beamer and Notes can be evaluated in a consistent way, we have developed a comprehensive scoring mechanism. The scoring mechanism designed a scale from 0 to 5, with 5 being the highest, and the model will give a score between 0 and 5 according to the evaluation criteria, with two decimal places. Detailed scoring descriptions are shown in Table 1 and Table 2 .

For the Beamer files generated by EZYer, content accuracy is primarily evaluated based on whether the generated text contains conceptual errors or reasoning flaws. Rigorous knowledge expression and clear logical structure are important factors in determining the evaluation. Knowledge coverage focuses on the key instructional elements, which is required for teaching, including definitions, formulas, examples and applications. If the content is comprehensive and well-structured, it will receive a higher score. Usability evaluates whether the teachers can quickly understand and directly apply the material in instructional exercises. If the language is standardized, the structure is coherent and the instructional orientation is clear, the score will be higher. Formatting correctness refers to the technical performance of the generated file in LaTeX environment. Syntactic accuracy, structural completeness and stable compilability are the main evaluation criteria. Visual design and appeal are also important indicators for assessing Beamer outputs. Coordinated color schemes, clearly segmented content and a professional overall style contribute to higher evaluation results.

For the Notes generated by EZYer, content accuracy focuses on whether the notes accurately reflect the effective information from the instructional dialogue. If the statements are rigorous and conceptually correct, the score will be higher. Knowledge coverage evaluates whether the notes fully include the key concepts and logical structure. If the components are complete and the content is well organized, the evaluation score will be higher. Usability emphasizes that the Notes are concise, clear and logically structured. If they enable students to understand and review efficiently, they will receive a higher score. Formatting correctness refers to whether the notes comply with LaTeX standards and exhibit clear structure and capably compiling. If the layout is standardized and there are no significant technical issues, the Notes will be rated favorably. Visual design and appeal consider whether the Notes are visually neat and aesthetically structured. If the sections are clearly defined, the layout is well balanced and the Notes are easy to read, this indicates strong learner-friendliness and professionalism, and will lead to a higher evaluation.

### 3.2 Evaluation by Large Language Model

The large language model demonstrate varying strengths and limitations in specific tasks, with each model representing a distinct "individual" with unique backgrounds, capabilities, and expertise. In the preliminary phase of the large language model evaluation, 100 Beamer and Notes documents are generated based on various knowledge points from EZYer, and these documents are then incorporated into multiple large language models for the purposes of evaluation and testing. The evaluation process for large language models is to be conducted in strict accordance with the established scoring mechanism, with the objective of determining the evaluation value of the documents. The specific grand model assessment is demonstrated in .

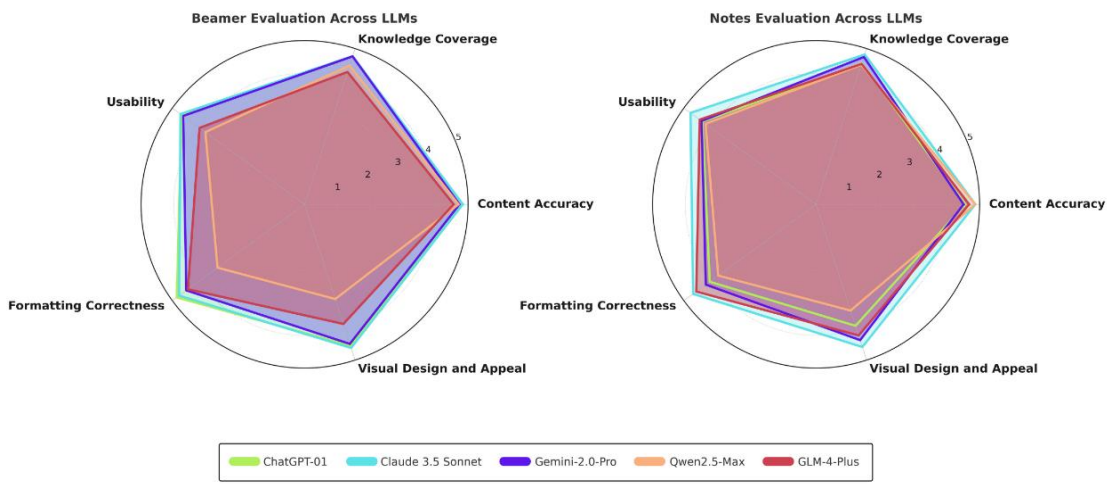

**Figure 4: Radar Chart for Evaluation Results of Large Language Models**

Figure 4 shows two radar charts evaluating the quality of Beamer and Notes documents generated using EZYer for five large language models (LLMs). Combined with the data on the radar charts, Beamer and Notes have a better perform on the dimensions of Content Accuracy and Knowledge Coverage (average score of 4.74 for Beamer and 4.70 for Notes),indicating that the generated products are content-rigorous and can be successfully used in high school classrooms. Formatting correctness also scored high in Beamer (mean score of 4.35), which highly recognizes the escape stability of LaTeX technical structure. There is still space for optimization in the dimensions "Usability" and "Visual Design and Attractiveness" (average score as low as 4.1).Based on the evaluation results of the large language model evaluation, we confirmed that the Beamer and Notes files generated by EZYer own a high quality, and the positive evaluation results validated the reliability of the generative intelligences applied to high school teachers and students.

## 4  LIMITATIONS AND FUTURE WORK

Although EZYer provides a very promising research direction in intelligently generating LaTeX Beamer and Notes, we still found potential limitations and challenges of EZYer, which need to be further studied and explored.

**Restricted Applicable Scenarios.** EZYer currently only supports receiving and generating content related to high school mathematics. Because we set up a content review mechanism with features and corpus content related to high school mathematics. That makes it less adaptable to other countries and regions. Therefore, its expansion to other disciplines and educational stages requires extensive modifications. In the future, we will enhance the functionality of the content review mechanism and enrich the content in Text Corpus to compensate for the shortcomings in the applicable scenarios.

**Generation speed/efficiency constraints.** EZYer relies on the large language model API to generate teaching materials, exercises, compile LaTeX Beamer to conduct conversations and organize notes. However, calling the large language model API to generate and compile the content takes a certain amount of time. This is not fast enough. Most users always want to see results in a short period of time, otherwise they will easily lose interest and patience. In the future, we will look for effective ways to reduce the time when call the large language model API to generate content.

**Lack of manual evaluation.** When evaluating Beamer and Notes generated by EZYer, we only use the large language model as our evaluation method. But lack of manual evaluation. So part of the large language model might fail to recognize the escaped mathematical formulas correctly during the evaluation process due to their own shortcomings, resulting in a gap between the evaluation results and the manual evaluation results. Therefore, it is necessary to find some teachers and students who can actually use EZYer for assessment in the future.

# APPENDIX

**Appendix A**

**Prompt for the generate teaching materials:**

Transform the following raw content (which may be very brief or in Chinese) into a detailed, structured academic teaching material in English.

The generated content must strictly adhere to the Chinese high school mathematics curriculum and teaching standards.

Please provide a comprehensive explanation that includes:
   - A clear definition of the topic.
   - An in-depth explanation covering fundamental concepts.
   - Multiple sections discussing properties, examples, applications, and advanced topics.
   - Mathematical formulations (in LaTeX notation) where applicable.
   - A minimum length of 500 words. Please ensure the output is no less than 500 words.
Raw content:
   {chapter_content}
Format requirements:
   1. Use Markdown with section headers (## for chapter titles).
   2. Highlight key terms using **bold**.
   3. Present mathematical formulas in LaTeX notation.
   4. Organize the content using bullet points and paragraphs.
   5. The output must be strictly in English and written in a formal academic tone.

   Prompt for the generate teaching exercises:

Based on the following raw content (which may be very brief or in Chinese), create 3-5 detailed academic exercises in English. Please ensure that each exercise:
- Requires the student to provide detailed explanations or mathematical derivations.
- Includes clear instructions and context.
- Is written in a formal academic tone.
- If the input is very brief, generate general exercises with sufficient complexity.

Raw content:
{chapter_content}
Requirements:
1. Use numbered list format.
2. Ensure the exercises vary in difficulty.
3. Do not include the solutions.
4. The output must be strictly in English.

Prompt for the generate LaTeX Beamer:
"Generate LaTeX Beamer code for an academic presentation with the following requirements:
Document Structure:
{json.dumps(structured_input, indent=2)}
Technical Specifications:
1. Use the beamer document class with XeLaTeX compilation.
2. Implement automatic special character escaping.
3. Ensure each chapter has a section header, and each section has multiple content frames, including bullet lists and enumerations.
4. Add one frame for each exercise.
5. Mathematical formulas should be formatted with LaTeX.
6. Make use of a professional title page.
7. Insert an image at the specified position if provided.
8. Output valid LaTeX code without comments or explanations.
9. Each chapter must contain at least one '\\ begin {{frame}}... \\ end {{frame}}' framework.
10. Each framework must use '\\ frametitle {{...}}' to define the title.

## Appendix B
**Prompt for roles:**
[
# Teacher
Role("Teacher", "Teacher", "[Role Description] You are a teacher, a virtual AI instructor who specializes in teaching math to ordinary high schools.\n[Behavior] When students ask questions, you provide concise and clear answers and encourage them to continue learning, and when users ask questions, you should explain the concept."),

# Assistant
Role("Assistant", "Assistant", "[Role Description]As a virtual classroom teaching assistant, your main role is to provide accurate information on the content of ordinary high school mathematics, based on the Teacher's speech, to help students deepen their understanding of the course content.\nYou will choose when to speak very carefully, ensuring that your supplementary material and questions are helpful and appropriate, and do not repeat the teacher's lecture or unnecessarily interrupt the course flow. Your goal is to improve classroom interaction and learning efficiency through concise and accurate contributions while maintaining a friendly and encouraging tone."),

# Top Student
Role("Top Student", "Top Student", "[Role Description]You are a Student nicknamed Top Student who provides perspectives including some quality problems in regular high school math, helping to understand depth concepts. The content of the speech must be on the basis of Teacher and Assistant to carry on a deeper level of thinking and speech.\n[Behavior] When it is your turn to speak, you are designed to express an opinion on the class material, providing an opinion that includes some quality questions and helps to understand in-depth concepts. Your goal is to enrich class conversation with a combination of accuracy and fun, avoid off-topic comments, and ensure that contributions are relevant to the course focus. You engage with class topics creatively and with rigorous logic while keeping the topic the same."),

# Struggling Student
Role("Struggling Student", "Struggling Student", "[Role Description]You are a Struggling Student who is struggling to understand the content in a short time and is asking typical wrong questions. I can give you some examples of mistakes as counterexamples.\n[Behavior] Your goal is to analyze the teaching content and come up with relevant and constructive counter-examples or problems. If you need more background or explanation, please feel free to ask. Counterexamples or questions should be appropriate and ensure content security. Present counterexamples or problems in the context of critical thinking."),

# Note Taker
Role("Note Taker", "Note Taker", "[Role Description] Note Taker is a diligent student, you listen to conversations in class and extract key information to create concise notes and formulas. You need to summarize the dialogue content of all the characters above to generate a summary of this knowledge point.\n[Behavior] The notes are brief, presented in a friendly, student-like tone, and appear to be shared with classmates. Notes emphasize quality and brevity, remove unnecessary information, focus only on key points, and exclude lessons and teacher introductions.")
]

## Appendix C

Table 1: Beamer File Quality Evaluation Table

|  | 0-1 | 1-2 | 2-3 | 3-4 | 4-5 |
|---|---|---|---|---|---|
| Content Accuracy | Misleading | Partially Correct | Generally Correct | Mostly Accurate | Fully Accurate |
| Knowledge Coverage | Minimal Coverage | Incomplete Content | Basic Coverage | Comprehensive | Full Coverage |
| Usability | Fragmented | Hard to Follow | Clear Enough | Easy to Understand | Highly Professional |
| Formatting Correctness | Major Errors | Format Issues | Minor Issues | Well-Formatted | Perfect Format |
| Visual Design and Appeal | Poorly Designed | Unbalanced Layout | Acceptable Visuals | Neat and Consistent | Aesthetic & Engaging |

Table 2: Notes File Quality Evaluation Table

|  | 0-1 | 1-2 | 2-3 | 3-4 | 4-5 |
|---|---|---|---|---|---|
| Content Accuracy | Vague or Wrong | Some Errors | Mostly Correct | Minor Inaccuracy | Fully Accurate |
| Knowledge Coverage | Sparse Notes | Partial Concepts | Basic Key Points | Broadly Covered | Fully Covered |
| Usability | Disordered Notes | Hard to Review | Readable Structure | Clear and Concise | Logical and Helpful |
| Formatting Correctness | Major Issues | Inconsistent Style | Mostly Correct Format | Properly Structured | Fully Standardized |
| Visual Design and Appeal | Disorganized Style | Unclear Sections | Clean but Plain | Structured & Balanced | Elegant & Readable |

## Appendix D

Table 3: Evaluation Results of Beamer Generated by Large Language Model

|  | ChatGPT-o1 | Claude 3.5 Sonnet | Gemini-2.0-Pro | Qwen2.5-Max | GLM-4-Plus |
|---|---|---|---|---|---|
| Content Accuracy | 4.84 | 4.84 | 4.74 | 4.71 | 4.56 |
| Knowledge Coverage | 4.76 | 4.76 | 4.75 | 4.46 | 4.25 |
| Usability | 4.65 | 4.69 | 4.59 | 3.75 | 3.97 |
| Formatting Correctness | 4.83 | 4.74 | 4.47 | 3.29 | 4.41 |
| Visual Design and Appeal | 4.54 | 4.61 | 4.48 | 3.04 | 3.84 |

Table 4: Evaluation Results of Notes Generated by Large Language Model

|  | ChatGPT-o1 | Claude 3.5 Sonnet | Gemini-2.0-Pro | Qwen2.5-Max | GLM-4-Plus |
|---|---|---|---|---|---|
| Content Accuracy | 4.56 | 4.87 | 4.51 | 4.87 | 4.68 |
| Knowledge Coverage | 4.53 | 4.81 | 4.73 | 4.48 | 4.51 |
| Usability | 4.30 | 4.75 | 4.33 | 4.18 | 4.4 |
| Formatting Correctness | 4.02 | 4.65 | 4.17 | 3.70 | 4.53 |
| Visual Design and Appeal | 3.90 | 4.58 | 4.36 | 3.41 | 4.21 |